\documentclass[published]{nst}

\usepackage{makecell}
\usepackage{siunitx}
\usepackage{subfigure,dcolumn}
\usepackage[T2A,T1]{fontenc}
\usepackage[russian,english]{babel}
\usepackage{bm}

\usepackage{listings}
\makeatletter

\newcommand{\Rmnum}[1]{\expandafter\@slowromancap\romannumeral #1@}
\makeatother

\begin{document}

\title{Layered reconstruction framework for longitudinal segmented electromagnetic calorimeter}\thanks{}

\author{Jia-Le Fei}
\affiliation{The Institute for Advanced Studies, Wuhan University, Wuhan 430072, China}
\author{Ao Yuan}
\affiliation{The Institute for Advanced Studies, Wuhan University, Wuhan 430072, China}
\author{Ke Wei}
\affiliation{School of Physics and Technology, Wuhan University, Wuhan 430072, China}
\author{Liang Sun}
\email[Corresponding author, ]{E-mail address: liang.sun@cern.ch}
\affiliation{School of Physics and Technology, Wuhan University, Wuhan 430072, China}
\author{Ji-Ke Wang}
\email[Corresponding author, ]{E-mail address: Jike.Wang@cern.ch}
\affiliation{The Institute for Advanced Studies, Wuhan University, Wuhan 430072, China}

\begin{abstract}
In future high-energy physics experiments, the electromagnetic calorimeter (ECAL) should operate with an exceptionally high luminosity. An ECAL featuring a layered readout in the longitudinal direction and precise time-stamped information offers a multidimensional view, thereby enriching our understanding of the showering process of electromagnetic particles in high-luminosity environments. This was used as the baseline design for several new experiments, including the planned upgrades of the current running experiments. Reconstructing and matching multidimensional information across different layers poses new challenges for the effective utilization of layered data. This study introduced a novel layered reconstruction framework for ECAL with a layered readout information structure and developed a corresponding layered clustering algorithm. This expands the concept of clusters from a plane to multiple layers. Additionally, this study presents the corresponding layered cluster correction methods, investigates the transverse shower profile utilized for overlapping cluster splitting, and develops a layered merged $\pi^0$ reconstruction algorithm based on this framework. By incorporating energy and time information into 3-dimensions, this framework provides a suitable software platform for preliminary research on longitudinally segmented ECAL and new perspectives in physics analyses.

Furthermore, using the PicoCal in LHCb Upgrade \Rmnum{2} as a concrete example, the performance of the framework was preliminarily evaluated using single photons and $\pi^{0}$ particles from the neutral $B^0$ meson decay $B^0\rightarrow\pi^+\pi^-\pi^0$ as benchmarks. The results demonstrate that, compared to the unlayered framework, utilizing this framework for longitudinally segmented ECAL significantly enhances the position resolution and the ability to split overlapping clusters, thereby improving the reconstruction resolution and efficiency for photons and $\pi^0$s.
\end{abstract}


\maketitle

\section{Introduction}\label{sec.I}

An electromagnetic calorimeter (ECAL) is a detector used to measure the energy and momentum of high-energy electromagnetic particles (e.g., electrons and photons). In particle physics, ECALs can measure the electromagnetic showering of particles within it and determine the energy and momentum of electromagnetic particles based on energy deposition, providing important information for understanding the properties and interactions of elementary particles.

A high-performance ECAL is crucial for the precise detection of high-energy physical phenomena and has been validated in many experiments\cite{nst1,nst2,EM_in_NICA,ALICE:2022qhn}. For instance, in the context of the LHCb experiment, during Runs 1 and 2 of the Large Hadron Collider (LHC), approximately 33$\%$ of the decay products of heavy-flavor particles are neutral particles, which decay into photons, such as $\pi^0$\cite{Lindner:2866493}. These photons exhibit a broad energy spectrum, ranging from a few gigaelectronvolts to several hundred gigaelectronvolts \cite{BOETTCHER2019251}. Owing to the outstanding performance of ECAL in the LHCb\cite{LHCb_detector_performance,LHCb_DP_2020_001}, Runs 1 and 2  yielded many impactful studies involving photons, $\pi^0$s, and electrons. This research explored photon polarization in the $b \rightarrow s\gamma$ process\cite{PhysRevLett.112.161801}, radiative $B_s^0$ decay \cite{Photon_Polarization_Bs0}, and other related studies. In addition, quantification of CP violations in decays, such as $B^+ \rightarrow K^+\pi^0$ \cite{PhysRevLett.126.091802} and $D^0 \rightarrow \pi^+ \pi^- \pi^0$ decays\cite{CPV_D02PiPiPi0}, was conducted. Furthermore, intriguing investigations into lepton universality through the reconstruction of $b \rightarrow s\ell^+\ell^-$ transitions have been conducted \cite{test_lepton_universality}\cite{test_lepton_universality2}. This research also includes rare decay searches, such as $B_s^0\rightarrow \mu^+\mu^-\gamma$\cite{mumuGamma}.

In pursuit of new physics at the LHC, a high-luminosity LHC (HL-LHC) project was proposed to enhance the accumulation of collision data within a shorter timeframe \cite{CERN-LHCC-2017-003, LHCBFTDR, Aaij:2636441}. Operating with an instantaneous luminosity of $10^{34}\ \mathrm{cm^{-2}s^{-1}}$, the HL-LHC aims to achieve impressive integrated luminosity over its operational lifetime \cite{LHCBFTDR, Efthymiopoulos:2319258}. However, such a high-luminosity environment presents challenges, such as increased detector occupancy, vertex pile-ups, and radiation resistance concerns. Many current detectors at the LHC are no longer adequate for operation under HL-LHC conditions, and long-term radiation damage necessitates urgent replacements. To ensure optimal detector performance in high-luminosity environments and enable exploration of new physics, several experiments are planning detector upgrades \cite{CMS_U2, ATLAS_U2, ALICE_U2,nst3,nst4}.

The showering process of electromagnetic particles in an ECAL is influenced by factors such as particle type and incident energy, which in turn affect the distribution of the deposited energy \cite{livan2019}. An ECAL equipped with multiple readout channels spanning both the transverse and longitudinal directions facilitates the capture of time, energy, and other readout data in dual dimensions. This capability significantly enhances our comprehensive understanding of the physical processes within the ECAL across multiple dimensions, ultimately leading to improved accuracy in reconstructing the energy and momentum of the particles. Therefore, a longitudinally segmented ECAL has been considered a baseline design for upgrades in many experiments.

The development of event reconstruction algorithms that fully exploit the energy, position, and time information acquired from longitudinally segmented ECAL will enhance the precision of energy-momentum reconstruction, cluster splitting, and particle identification, thereby facilitating the achievement of defined physical objectives. However, this also introduces new challenges in the reconstruction, matching, and efficient utilization of longitudinal-layer information. To leverage the advantages of layered readouts in forthcoming high-energy physics experiments, such as CMS and ALICE \cite{CMS_Alg_e, CMS_Alg_e_gamma, CMS_Alg_gamma, ALICE_Rec,CLUE,TICL}, diverse software frameworks and reconstruction algorithms have been devised and customized to effectively utilize and store layered information.

Building on this background, this study presents a comprehensive software framework tailored for a longitudinally segmented ECAL, along with the development of a layered clustering algorithm and cluster correction workflow. The layered reconstruction framework outlined in this study merges data from all the layers to identify potential $Cluster^{3D}$ candidates. Subsequently, a seed is identified in each layer from the readout cells of the $Cluster^{3D}$ candidates. Utilizing these seeds as focal points, $Cluster^{2D}$s in all readout layers are reconstructed (the definitions of $Cluster^{3D}$ and $Cluster^{2D}$ are elaborated on in subsequent sections). Finally, the performance is compared with that of an unlayered-reconstruction algorithm that combines the information from the corresponding readout units in each layer based on the single-photon resolution, as well as the reconstruction resolution and efficiency for $\pi^{0}$ from $B^0\rightarrow\pi^+\pi^-\pi^0$. This comparison leveraged the 2023 baseline setup of ECAL in the LHCb Upgrade \Rmnum{2} (PicoCal)\cite{Kholodenko:2836693,Scoping_Document}.

\section{Software framework and data structures}\label{sec.II}
\begin{figure*}[!bht]
\includegraphics
  [width=0.95\hsize]
  {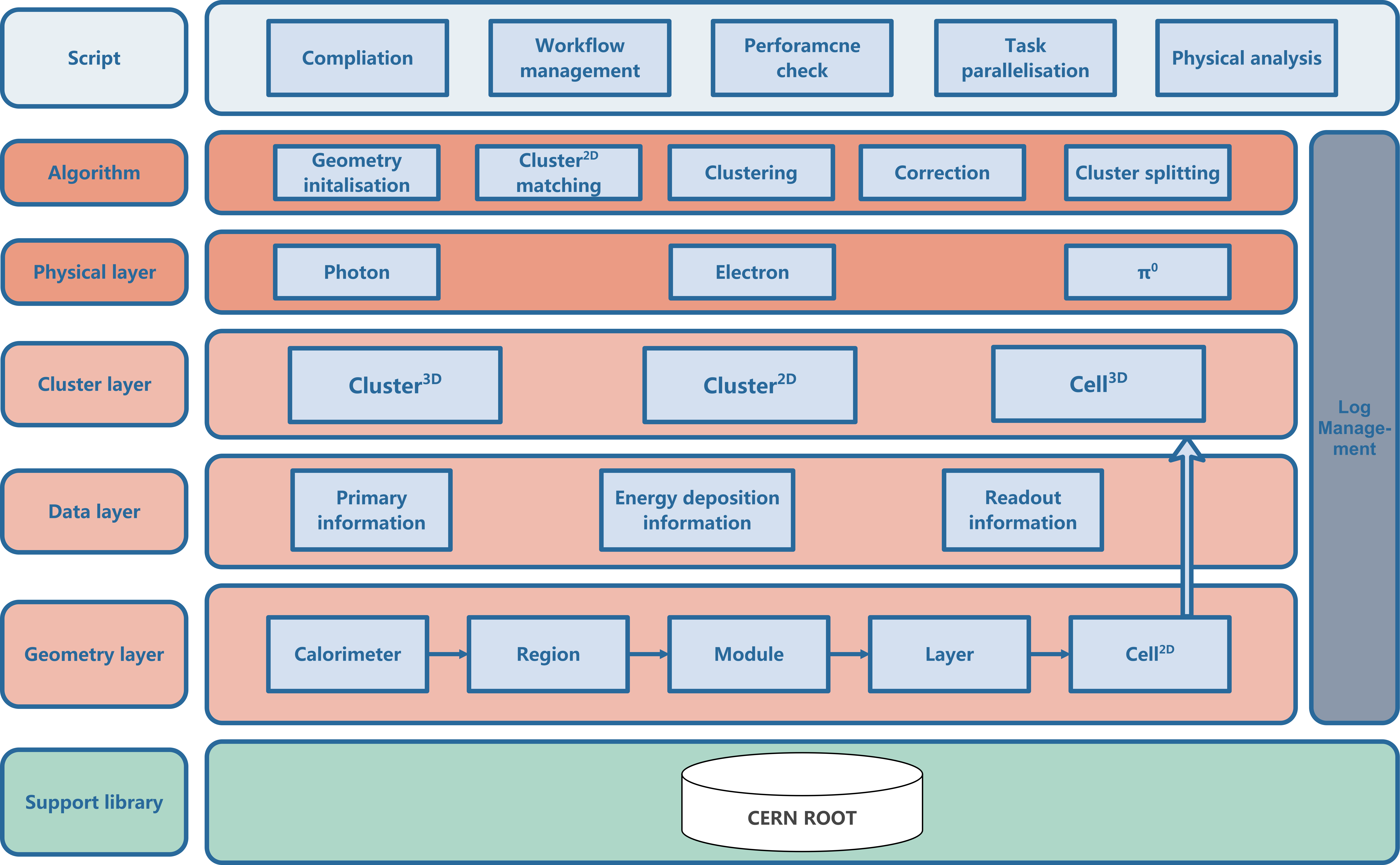}
\caption{Software framework.}
\label{fig:Reconstruction_Framework}
\end{figure*}

The layered reconstruction framework is shown in Fig.~\ref{fig:Reconstruction_Framework}. The modeling of the detector geometry is the cornerstone of the framework. To adapt as many longitudinally segmented ECAL structures as possible, the following data structures were constructed to describe and carry the geometrical information of the ECAL in this framework.
\begin{description}

\item[Calorimeter] It represents the ECAL and also stores the absolute coordinates and size of the entire ECAL in space.
\item[Region] It is a virtual geometry containing a series of \textbf{module}s with identical detector structure, material, and installation angle. It was used to store the calibration and cluster correction parameters for this series of \textbf{ modules }.
\item[Module] It represents the minimum installation unit and stores the absolute coordinates, size and installation angle of itself. 
\item[Layer] It is the key geometric data structure in this framework and physically represents a longitudinal segment of the \textbf{module}. It stores all $\textbf{Cell}^\textbf{2D}$s located in a segmented readout layer in a \textbf{module}.
\item[$\textbf{Cell}^\textbf{2D}$] It represents the minimum readout channels in modules and contains information regarding the mounting position of the readout cell, signal, and timestamp.
\end{description}

Based on the aforementioned basic geometric structure, the following cluster data structures, containing two or more $Cell^{2D}$s, were constructed, which correspond to the cluster layer in the software framework. The specific construction method for the data structures is described in Sec.~\ref{sec:Clustering}.

\begin{description}

\item[$\mathbf{Cell^{3D}}$] It consists of multiple $Cell^{2D}$s in the longitudinal direction.
\item[$\mathbf{Cluster^{2D}}$] It consists of multiple $Cell^{2D}$s at the same layer.
\item[$\mathbf{Cluster^{3D}}$] It consists of a $Cluster^{2D}$ and multiple $Cell^{2D}$s in each layer and would be considered as a candidate for photons, electrons, or merged $\pi^0$s.
\end{description}

\section{Reconstruction}\label{sec.III}
Based on the layered reconstruction framework, this section describes the reconstruction algorithm for $Cluster^{3D}$, which is a candidate for electromagnetic particles ($\gamma/e$).  The methodology for reconstructing a merged $\pi^0$ by splitting a single $Cluster^{3D}$ is also discussed. In addition, it outlines the process for correcting various parameters of $Cluster^{3D}$, including the energy, position, and time.

\subsection{Layered clustering} \label{sec:Clustering}
The algorithm outlined in this section details the methodology used to finalize the construction of  $Cluster^{3D}$s and provides particle reconstruction information at each readout layer in the form of $Cluster^{2D}$s. A flowchart of the layered clustering algorithm is shown in Fig.~\ref{fig:CluFlow2}d. First, the algorithm searches for $Seed^{3D}$s on a single layer, which includes a realistic readout layer with a smaller transverse showering width, as well as a virtual single-layer calorimeter constructed by merging information from all readout layers and constructing temporary $Cluster^{3D}$s in the virtual single-layer calorimeter from $Seed^{3D}$s. This approach was motivated by two primary considerations. 

The first and most important point is that constructing $Cluster^{2D}$ in each layer and performing layer-by-layer matching consumes a significant amount of computation time. First, performing 2-dimensional clustering in a single layer ensures efficient online triggering. 

Second, in layers with narrower transverse cluster development, there is a chance of discovering more non-overlapping clusters. However, owing to the narrower transverse cluster development, the energy deposition is lower. Owing to sampling fluctuations, some particles may not have formed effective seeds in this layer. Therefore, searching for $Seed^{3D}$s in both $Cell^{3D}$s and $Cell^{2D}$s in layers with narrower cluster development will help achieve a balance between cluster separation and reconstruction efficiency.

Subsequently, based on the temporary $Cluster^{3D}$s, we construct $Cluster^{2D}$s located in different layers to obtain the final $Cluster^{3D}$s. Information from different layers was initially integrated by constructing a $Cell^{3D}$. The details of each step of the layered clustering algorithm are as follows:

\begin{figure*}[!bht]
\includegraphics
  [width=0.95\hsize]
  {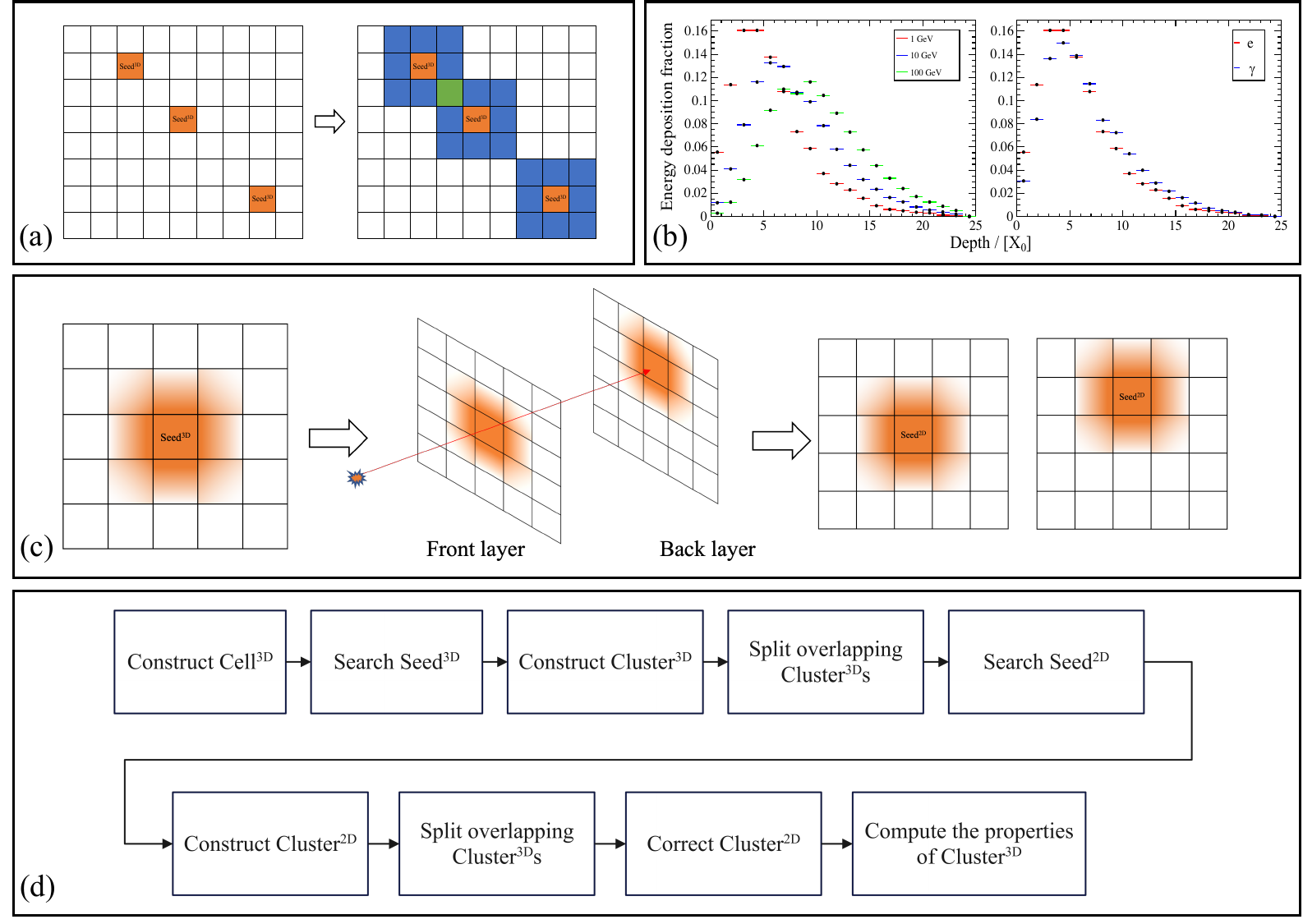}
\caption{(a): Searching $Seed^{3D}$s and constructing temporary $Cluster^{3D}$s centered on $Seed^{3D}$s, where the orange boxes represent $Seed^{3D}$s, blue boxes represent $Cell^{3D}$s, and green box represents $Cell^{3D}$ shared by two $Cluster^{3D}$s. (b): The energy deposition in the longitudinal direction of ECAL, where the left plot shows electrons with different energies and the right plot shows the electron and photon with 1 GeV energy. (c): Searching for $Seed^{2D}$ in the $Cell^{2D}$ of the temporary $Cluster^{3D}$ and constructing new $Cluster^{2D}$ in each layer. (d): Workflow chart for layered reconstruction.}
\label{fig:CluFlow2}
\end{figure*}

\subsubsection{Constructing $Cell^{3D}$}

To integrate the information in each readout layer, the information from the corresponding $Cell^{2D}$s in both layers is utilized to establish a novel data structure known as $Cell^{3D}$. Specifically, $Cell^{3D}$s were systematically constructed across the transverse section of the ECAL, with a radius equivalent to the module's Molière radius. Along the longitudinal direction of the ECAL, all $Cell^{2D}$s falling within the transverse span of $Cell^{3D}$ are amalgamated into $Cell^{3D}$, with each $Cell^{2D}$ being included only once. The energy of $Cell^{3D}$ is determined by the cumulative energy of the $Cell^{2D}$s it encompasses, and the position of $Cell^{3D}$ is defined as its geometric center.

\subsubsection{Searching $Seed^{3D}$ and constructing temporary $Cluster^{3D}$}
When an electromagnetic particle strikes the ECAL, it radiates energy outward from the impact point, typically resulting in the formation of $Cell^{2D}$ and $Cell^{3D}$ with the highest local energy deposition near the impact point. Therefore, as illustrated in Fig.~\ref{fig:CluFlow2}a, the initial step in clustering is to iterate through all $Cell^{3D}$s and $Cell^{2D}$s in specific layers to identify the $Cell^{3D}$ and $Cell^{2D}$ layers that exhibit the highest local energy deposition. For the local maximum $Cell^{2D}$, the corresponding $Cell^{3D}$ is considered a $Seed^{3D}$. Simultaneously, the local maximum $Cell^{3D}$ is considered as $Seed^{3D}$. $seed$ denotes the initial step in the clustering process.

To prevent faker $Seed^{3D}$s during the seeding process, all identified $Seed^{3D}$ s must satisfy a transverse momentum cut, typically defined as greater than 50 MeV in this framework. The threshold value of this cut can be lowered if required to investigate phenomena related to soft photons or electrons. All $Seed^{3D}$s passing through the cut are labeled as final $Seed^{3D}$s and stored. 

As a result of transverse particle showering, the energy is not fully contained within $Seed^{3D}$. Hence, it is imperative to encompass a specific range of $Cell^{3D}$s around $Seed^{3D}$ to guarantee optimal coverage of all energy deposits from the particles. In this study, we followed the methods described in Ref.~\cite{LHCb-2001-123}, which uses a fixed-size window to incorporate $Cell^{3D}$s.
The specific procedure is as follows: Centered around the $Seed^{3D}$s, the $Cluster^{3D}$s are formed by incorporating all $Cell^{3D}$s within a $3\times 3$ window around the $Seed^{3D}$s, as illustrated in Fig.~\ref{fig:CluFlow2}a. Additionally, the $Cell^{2D}$s from all layers encompassed by the $Cell^{3D}$s are also included as members of the $Cluster^{3D}$s. However, when the $Seed^{3D}$ s are located at the boundary of the region, special treatment is required because of the different types of $Cell^{3D}$s in different regions. Around $Seed^{3D}$, within a radius of 1.5 times the size of $Seed^{3D}$, $Cell^{3D}$s belonging to other regions are also included. In this study, a detailed description of the process is not provided. This process may lead to $Cluster^{3D}$ at the boundary containing several $Cell^{3D}$s, which may be greater or less than nine, and its shape may be irregular. This requires separate handling of $Cluster^{3D}$ at the boundary in the $Cluster^{3D}$ correction process described later.

In the subsequent steps, further layered modifications were made to the $Cell^{2D}$s included in $Cluster^{3D}$. Therefore, $Cluster^{3D}$ obtained in this section is referred to as a temporary $Cluster^{3D}$.

\subsubsection{Searching $Seed^{2D}$ and reconstructing $Cluster^{2D}$}
Because of the incident angle of the particles and the rotation of certain modules, as the shower evolves longitudinally, the energy centroid of the shower varies across different layers. This results in $Cell^{2D}$  with the highest local energy in each layer, not always being encompassed within $Seed^{3D}$, as shown in Fig.~\ref{fig:CluFlow2}c.

To identify the $Seed^{2D}$ in each layer, we iterate through the $Cell^{2D}$s of each layer in the temporary $Cluster^{3D}$ and select the $Cell^{2D}$ with the highest energy as the $Seed^{2D}$ of that layer. Moreover, the overlap of $Cluster^{2D}$s results in an increased accumulation of energy in shared $Cell^{2D}$s, which may cause the energy of a shared $Cell^{2D}$ to exceed that of a $Seed^{2D}$, leading to the misidentification of a shared $Cell^{2D}$ as a $Seed^{2D}$. Therefore, before searching for $Seed^{2D}$, energy splitting must be performed on all overlapping $Cluster^{2D}$s, as shown in Fig.~\ref{fig:CluFlow2}d, and is discussed in detail in Sec.~\ref{Split Energy}.

Subsequently, $Cluster^{2D}$s are formed with $Seed^{2D}$s as the center, and all $Cell^{2D}$s within the Molière radius of the module centered on $Seed^{2D}$s in the same layer are included in the $Cluster^{2D}$s. The raw energy, position, and timestamp of the $Cluster^{2D}$s were computed using the following equations:

\begin{equation}\label{eq:Raw information}
\begin{aligned}
E_{raw} &= \sum_{i=1}^{n} E_i, \\
r_{raw} &= \frac{\sum_{i=1}^{n} E_i\times r_i}{E_{raw}}; \quad r = x\;or\;y, \\
t_{raw} &= t_{Seed^{2D}},
\end{aligned}
\end{equation}
where $n$ represents the number of $Cell^{2D}$s in $Cluster^{2D}$, and $r_i$ and $E_i$ represent the position and energy of $Cell^{2D}$s, respectively.

Finally, a new $Cluster^{3D}$ containing $Cluster^{2D}$s in all layers is constructed. The raw energy of $Cluster^{3D}$ is computed using the following formula:  
\begin{equation}\label{eq:Raw information}
E^{3D}_{raw} = \sum_{i=1}^{n} E_i, 
\end{equation}
where $E_i$ represents the energy of the $Cell^{2D}$s, and $n$ represents the number of $Cell^{2D}$s in all the layers incorporated in $Cluster^{3D}$.
As shown in Fig.~\ref{fig:CluFlow2}b, the energy deposition distribution of the incident particles depends on their type and energy of the incident particles. Therefore, unreasonable $Cluster^{3D}$s can be filtered based on the energy ratio of $Cluster^{2D}$s in each layer. The position information of $Cluster^{2D}$s is considered as a point of the incident particle momentum direction on the corresponding layer, with the time information serving as the timestamp for this coordinate point. Additionally, the energy information is regarded as the total deposited energy of the particles in that layer. Reconstructing and correcting the information of the particles along the momentum direction layer-by-layer helps improve the position resolution, which will be discussed in detail in Sec.~\ref{sec.IV}.

\begin{figure*}[!thb]
\includegraphics
  [width=0.78\hsize]
  {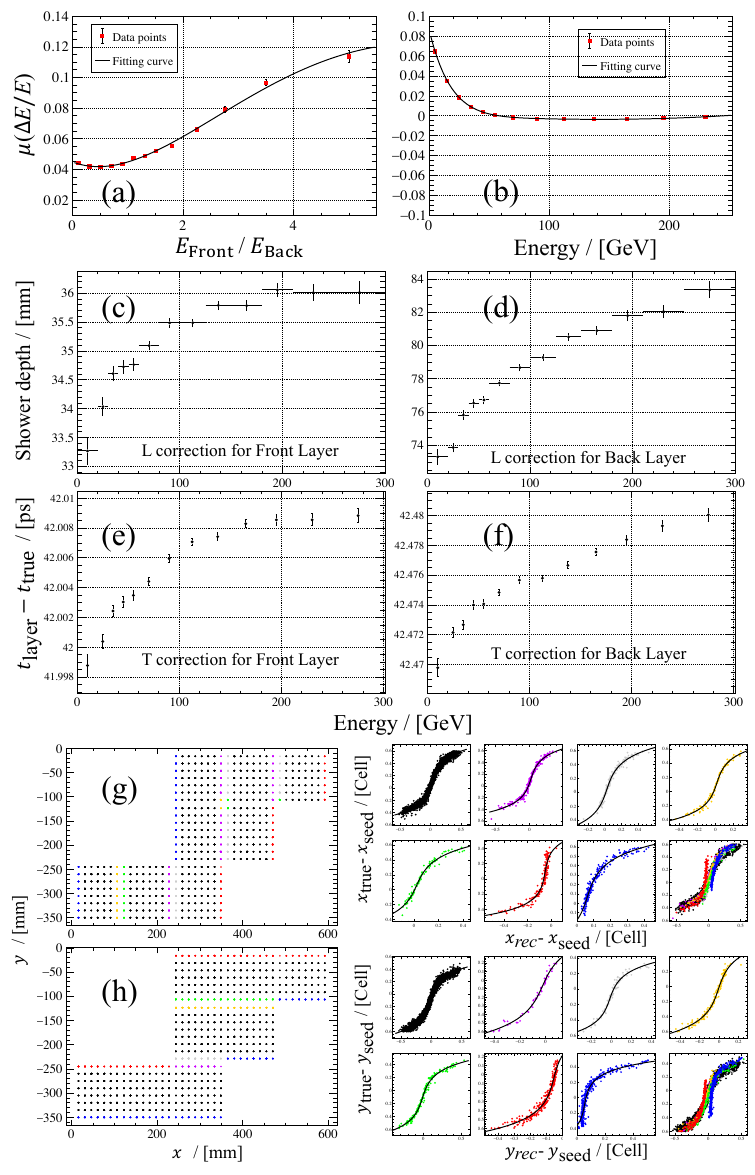}
\caption{(a, b): The energy correction. Here, $\Delta E$ is the difference between the true energy and the reconstructed energy, $E$ is the true energy, and $\mu$ represents the mean value which is derived from the Gaussian fitting. (c, d): The $L$ correction, where the shower depth is measured from the front surface of the ECAL. (e, f): The time correction points, where $t_{true}$ is the true time of the particle at the front face of the ECAL. (g, h): The $S$ correction.}
\label{fig:Correction}
\end{figure*}

\begin{figure*}[!bht]
\includegraphics
  [width=0.9\hsize]
  {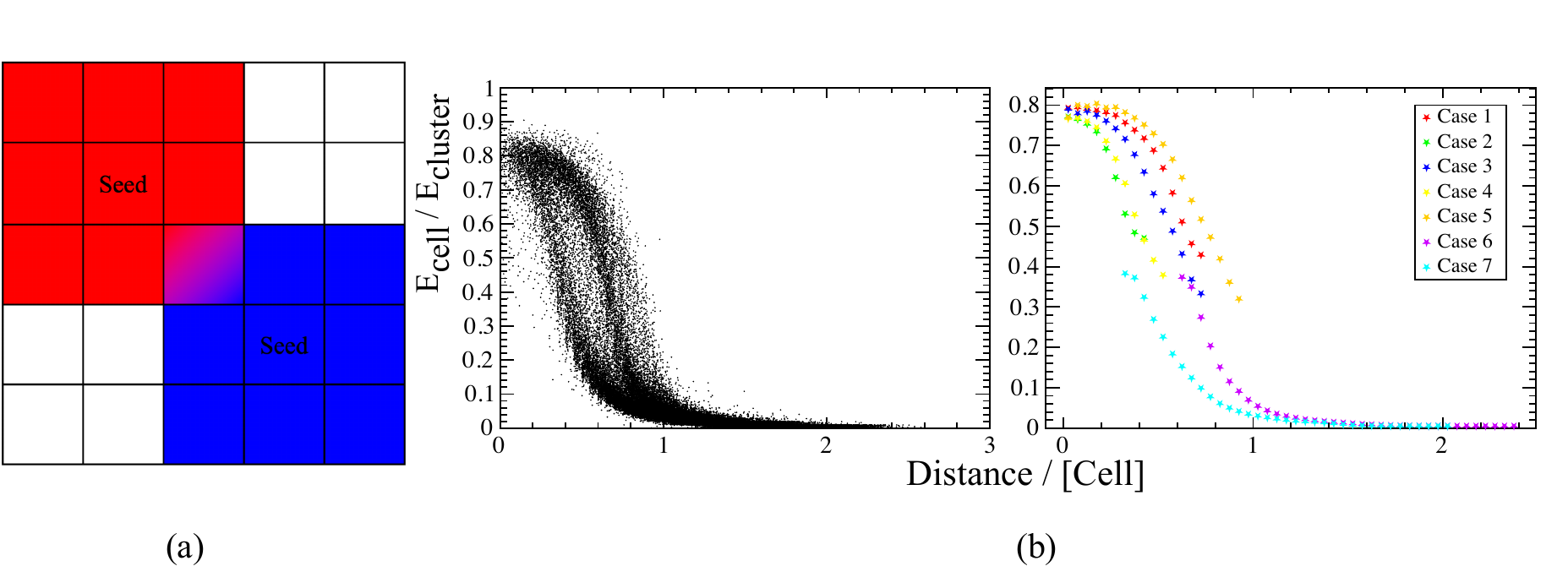}
\caption{(a): Two overlapping clusters shared the same cell. (b): The transverse shower profile, where the $y$ axis is the fraction of the energy of the $Cell^{2D}$s to the energy of the $Cluster^{2D}$s, and the $x$ axis is the distance between the center of the $Cell^{2D}$s and the center of the $Cluster^{2D}$s.}
\label{fig:ESplitting}
\end{figure*}

\subsection{$Cluster^{3D}$ correction}
The goal of the $Cluster^{3D}$ correction is to reduce the bias between the reconstructed and actual values, while also aiming for the minimal standard deviation of the reconstructed values. In the layered reconstruction framework, the position and timestamp of $Cluster^{3D}$ are calculated and corrected from the $Cluster^{2D}$ level. As described in Sec.~\ref{sec:Clustering}, the raw position and timestamp of $Cluster^{2D}$ can be calculated using information from the $Cell^{2D}$s in $Cluster^{2D}$. Subsequently, the raw position and timestamp information of $Cluster^{2D}$ are corrected, and the position and timestamp information of $Cluster^{3D}$ are calculated using the corrected $Cluster^{2D}$. The energy of $Cluster^{3D}$ was also corrected using the energy ratios of $Cluster^{2D}$s in different layers. To provide a more detailed illustration of the specifics of layered corrections, Fig.~\ref{fig:Correction} also displays some examples of the longitudinally segmented PicoCal in LHCb Upgrade \Rmnum{2}, where $Front$ and $Back$ represent the front and back layers of the abovementioned ECAL.

\subsubsection{Energy correction}
The objective of energy correction is to correct the energy of $Cluster^{3D}$ to match that of the incident particles. Errors in reconstructing the energy of an incident particle typically stem from the following sources:
\begin{description}
\item[Intrinsic error of ECAL readout cell] It comes from the response linearity of sensitive materials, thermal noise in electronic systems, sampling errors in analog to digital converters (ADC), etc.
\item[Calibration error of the readout cell] The shower development in the longitudinal direction is energy related, and the proportion of physical processes dominated in energy deposition changes at different stages of shower development. This leads to changes in the sampling fraction of the ECAL, which ultimately affects the calibration of the readout cell.
\item[Leakage of energy ] It is due to the incomplete deposition of particle energy in ECAL and the use of finite-sized windows during clustering.
\item[Fitting error of calibration and correction parameters] It is typically expressed as a constant term in the energy resolution.
\end{description}
In the layered framework of this study, the energy correction of $Cluster^{3D}$s was divided into two steps, as shown in Fig.~\ref{fig:Correction}a and Fig.~\ref{fig:Correction}b. The initial correction of the energy of $Cluster^{3D}$ was performed based on the energy ratios of $Cluster^{2D}$s in different layers. As shown in Fig.~\ref{fig:Correction}a, in the example based on PicoCal, we first correct the bias between the reconstructed $Cluster^{3D}$ energy and true energy based on the energy ratio of the front layer to the back layer. Subsequently, as illustrated in Fig.~\ref{fig:Correction}b, further correction of the energy of $Cluster^{3D}$s is conducted based on the total energy of $Cluster^{3D}$s, which reduces the bias in the low-energy region.

\subsubsection{Position and time correction}
When a particle passes through the ECAL, it showers and deposits energy along its momentum direction. The center of gravity of the deposited energy in each layer can be considered a point in the direction of the particle momentum. In the layered reconstruction framework, the position of $Cluster^{2D}$ is regarded as the reconstructed position of the center of gravity. Furthermore, $Cluster^{2D}$s also provide a timestamp, which is related to the hit time of the particle, for the layers to which they belong. For the position and time correction in this study, layered correction is applied to $Cluster^{2D}$s firstly, followed by the utilization of the corrected information from $Cluster^{2D}$s to calculate the information of $Cluster^{3D}$s.

The purpose of position correction is to correct the reconstructed position of $Cluster^{2D}$ as accurately as possible to the center of gravity of the energy deposited by the particles in each layer. 

For $Cluster^{2D}$, the $x/y$ position information was derived from the energy-weighted positions of the $Cell^{2D}$s. Based on the model described in Ref.~\cite{LHCb-2003-091}, we define $\Delta r_{rec}$ as the position of $Cluster^{2D}$ minus the position of $Seed^{2D}$, and $\Delta r_{true}$ as the position of the true transversal energy barycenter minus the position of $Seed^{2D}$, in each layer. This study explored the relationship between $\Delta r_{rec}$ and $\Delta r_{true}$. As shown in Fig.~\ref{fig:Correction}g and Fig.~\ref{fig:Correction}h, the shape of this relationship is like an $S$, so we also call the process of correcting the $x/y$ coordinates "S correction.” The shape of $S$ is affected by the location of $Seed^{2D}$. When $Seed^{2D}$ is positioned at the boundaries of the region, the introduction of a varying cell size $Cell^{2D}$ and the presence of installation gaps affect the impact on the $S$ shape. This results in a different $S$ shape for $Cluster^{2D}$s located at the boundary of the region compared with those inside the region, as illustrated in Fig.~\ref{fig:Correction}g and Fig.~\ref{fig:Correction}h.

The $z$ coordinate of the center of gravity of the deposition energy is typically used to evaluate the depth of the shower. Owing to the typically small granularity and thick layers of ECAL in the longitudinal direction, the direct use of the $z$-coordinate information of $Cell^{2D}$ to reconstruct the $z$-coordinate of the gravity center of the deposition energy in each layer introduces substantial uncertainty. Therefore, the reconstruction of the $z$-coordinate is typically performed based on the energies of the incident particles. As shown in Fig.~\ref{fig:Correction}c, and Fig.~\ref{fig:Correction}d, the shower depth (the difference between the $z$ coordinate of the shower and the $z$ coordinate of the front face of the module) is logarithmically related to the incident particle energy, because the pair production of electrons dominates the energy deposition\cite{Grupen_Shwartz_2008}. This is the rationale behind the term "L correction" for this correction step. By leveraging this correlation, we can deduce the $z$-coordinates of $Cluster^{2D}$ based on the energy of the incoming particles.

For the $Cluster^{2D}$s, the positions obtained by the above position correction were also projected onto the front surface of ECAL and then used in subsequent steps to calculate the position information of the $Cluster^{3D}$s.

In this study, the timestamp of the $Seed^{2D}$s were used as the timestamp for $Cluster^{2D}$s in the preliminary study. The purpose of time correction is to determine the time difference between the timestamps of $Cluster^{2D}$ in each layer and the moment when the particle reaches a specific reference plane. This time difference is employed to correct the timestamp of $Cluster^{2D}$ to an accurate time on a designated reference plane. The front surface of ECAL was used as the reference plane for time correction in this study. This time difference is related to the energy of the incident particle, and the correction results are shown in Fig.~\ref{fig:Correction}e and Fig.~\ref{fig:Correction}f. Furthermore, if there is a rotation of the module, it results in a longitudinal positional difference of $Cell^{2D}$s at different transverse positions in the same layer. This results in a time difference for the particles to reach different $Seed^{2D}$s on the same layer. Hence, it is necessary to compensate for this time difference based on the longitudinal position difference of $Seed^{2D}$s as follows:
\begin{equation}\label{eq:tSeed2D}
\begin{aligned}
t'_{Seed^{2D}}=(t_{Seed^{2D}}+\frac{z_{Seed^{2D}}-z_{layer}}{v_z}),
\end{aligned}
\end{equation}
where $t'_{Seed^{2D}}$ represents the compensated time of $Seed^{2D}$ and $v_z$ represents the velocity of the particle along the direction of the beam pipe.

After obtaining the corrected position and time information of $Cluster^{2D}$s, the time and position information of $Cluster^{3D}$s are obtained by weighting the time or position information of $Cluster^{2D}$s as follows:
\begin{equation}\label{eq:Weight$Cluster^{2D}$}
\begin{aligned}
V&= r\;or\;t, \\
W_i(V)&=\frac{1}{Res_i(V)^2}, \\
V_{Cluster^{3D}}&= \frac{\sum_{i=1}^{n}(V_{Cluster^{2D}_i}\times W_i(V))}{\sum_{i=1}^{n} W_i(V)},
\end{aligned}
\end{equation}

where $r$ represents the position, $t$ represents the time, $i$ represents the number of layers, and $Res_i$ represents the resolution of $r$ or $t$ in layer $i$, as shown in Fig.~\ref{fig:Resolution}b and Fig.~\ref{fig:Resolution}c.

\subsubsection{Splitting of overlapping clusters} \label{Split Energy}
When two $Cluster^{3D}$s overlap in an event, as shown in Fig.~\ref{fig:ESplitting}a, it is necessary to conduct energy splitting on the shared $Cell^{2D}$s to ensure accurate reconstruction of the energy and position of the $Cluster^{3D}$s. In the layered reconstruction framework, energy splitting of the overlapping $Cluster^{3D}$s is performed at the $Cluster^{2D}$ level. The general logic of the algorithm is as follows. First, determine if the two $Cluster^{3D}$s share any $Cell^{2D}$s. If they do, distribute the energy of the shared $Cell^{2D}$s between the respective $Cluster^{2D}$s. Upon completing the energy splitting of $Cell^{2D}$s, reevaluate the information of $Cluster^{2D}$s based on the updated energy of $Cell^{2D}$s, and remake the necessary corrections.

Currently, the energy splitting is determined by the transverse shower profile obtained from MC truth information. This study provides a layered description of the transverse shower profile at the $Cell^{2D}$ level. As shown in Fig.~\ref{fig:ESplitting}b, the transverse shower profile is represented by the distance of $Cell^{2D}$s from $Cluster^{2D}$s on the $x$-axis and the energy fraction of $Cell^{2D}$s relative to $Cluster^{2D}$s on the $y$-axis.

When $Cell^{2D}$ is shared by two $Cluster^{2D}$s, the distribution of $Cell^{2D}$ energy to each $Cluster^{2D}$ is evaluated in two steps. First, the energy fraction of the shared $Cell^{2D}$ corresponding to each $Cluster^{2D}$ is calculated based on the distance between $Cell^{2D}$ and $Cluster^{2D}$, as depicted in Fig.~\ref{fig:ESplitting}b. Second, the estimated energy from each of the two $Cluster^{2D}$s to $Cell^{2D}$ is computed using the fraction established in the first step and the energy of the $Cluster^{2D}$s. At this juncture, the total estimated energy from the two $Cluster^{2D}$s to the shared $Cell^{2D}$ exceeds that of the shared $Cell^{2D}$. The energy of the shared $Cell^{2D}$ is then distributed between the two $Cluster^{2D}$s, using the calculated estimated energy as the weighting factor. Subsequently, the energy and position of $Cluster^{2D}$ are recalculated, and the aforementioned procedures are repeated. Upon stabilizing the splitting weights of the shared $Cell^{2D}$ for the two $Cluster^{2D}$s, this iterative process completes and finalizes the energy splitting of the shared $Cell^{2D}$.

The precision of the energy fraction contributed by $Cell^{2D}$ to each $Cluster^{2D}$ relies heavily on the accuracy of the transverse shower profile. As depicted in Fig.~\ref{fig:ESplitting}b, even when the distances between $Cell^{2D}$s and their respective $Cluster^{2D}$s are identical, the energy contribution from $Cell^{2D}$s to $Cluster^{2D}$s can vary significantly and even display multiple peaks, particularly at distances of approximately 0.5 [Cell]. Therefore, it is essential to categorize the data points in the left-hand plot in Fig.~\ref{fig:ESplitting}b to achieve narrower fraction ranges corresponding to the same distance within a single category, as well as a singular peak in the energy fraction.

\begin{table*}[!htb]
\caption{The classification of transverse shower profile.}
\label{tab:ESplit_Classify}
\begin{tabular*}{16cm} {@{\extracolsep{\fill} } cccc}
\toprule
Case & Cell type& Hit position relative to Seed & Cell relative to Seed   \\
\midrule
1& Seed  & 1: In the positive direction of $p_{\mathrm{T}}$;  2: Near the edges of the Seed & /  \\
2& Seed  & 1: In the negative direction of $p_{\mathrm{T}}$;  2: Near the edges of the Seed & /  \\
3& Seed  & 1: In the positive direction of $p_{\mathrm{T}}$;  2: Near the corner of the Seed & /  \\
4& Seed  & 1: In the negative direction of $p_{\mathrm{T}}$;  2: Near the corner of the Seed & /  \\
5& Seed  & Near the corner closest to the direction of $p_{\mathrm{T}}$ of the Seed & /  \\
6& Cell  & / & In the positive direction of $p_{\mathrm{T}}$  \\
7& Cell  & / & In the negative direction of $p_{\mathrm{T}}$  \\
\bottomrule
\end{tabular*}
\end{table*}

The classification methods used in this study are listed in Table~\ref{tab:ESplit_Classify} and the classification results are shown in the right plot of Fig.~\ref{fig:ESplitting}b.  First, whether this $Cell^{2D}$ is a $Seed^{2D}$ should be determined. This is because the random scattering direction of the initial electron pairs affects which $Cell^{2D}$ near the hit point has the opportunity to receive more energy. A $Cell^{2D}$ deposited with the maximum energy was constructed as a $Seed^{2D}$ in the reconstruction process. Additionally, because the readout unit ($Cell^{2D}$) is not circular, the position of the hit point relative to the edges or corners of $Seed^{2D}$ affects the energy fraction of $Seed^{2D}$ when the distance to the hit point remains unchanged. Finally, the position of the hit point and the $Cell^{2D}$s relative to $Seed^{2D}$ in the positive or negative direction of the particle's transverse momentum also affect the energy fraction, which can be described as being relatively close to or far from the beam pipe in LHCb. In practical operations, the hit position of the particles in each layer is substituted by the position of $Cluster^{2D}$s.

\subsection{Merged $\pi^{0}$ reconstruction}
\begin{figure*}[!bht]
\includegraphics
  [width=0.92\hsize]
  {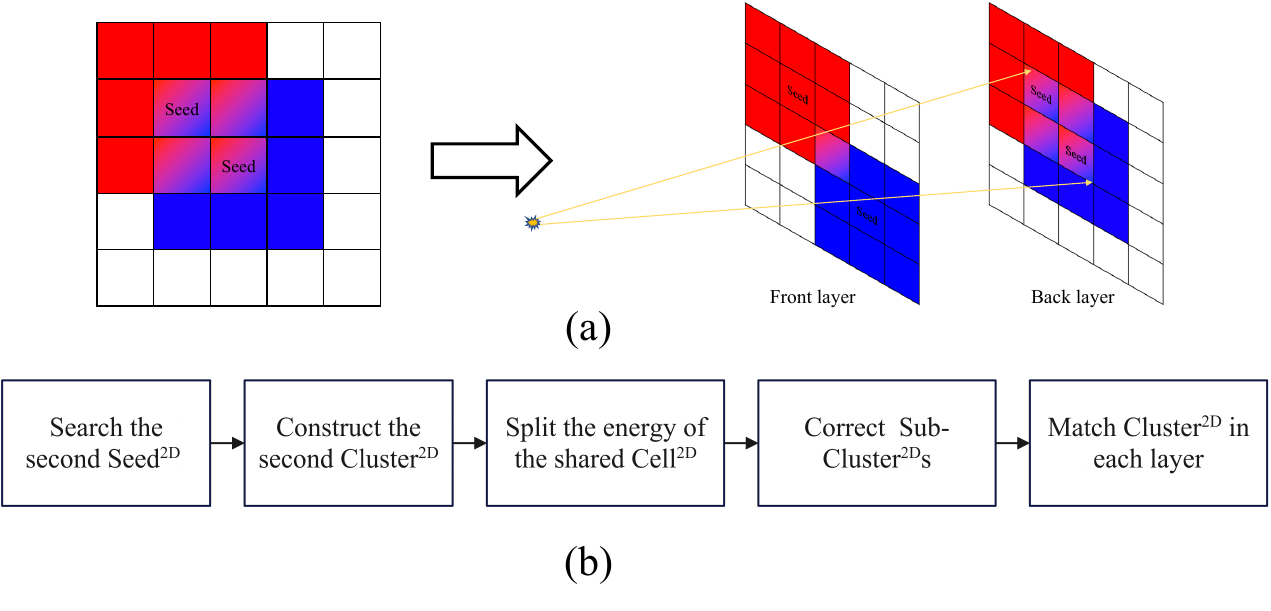}
\caption{(a): A possible $Cell^{3D}$s response and $Cell^{2D}$s response in each layer of a longitudinal segmented ECAL of a merged $\pi^{0}$. (b): Reconstruction flow chart for merged $\pi^{0}$.}
\label{fig:MergedPi0Rec}
\end{figure*}

When two photons produced by $\pi^{0}$ cannot be reconstructed individually as two $Cluster^{3D}$s because of the proximity of the hit points, $\pi^{0}$ is referred to as merged $\pi^{0}$. This section describes the reconstruction of the potential sub-$Cluster^{3D}$ pair, which is considered a candidate for the photon pair from the merged $\pi^{0}$, from a direct reconstruction of $Cluster^{3D}$. Subsequently, the sub-$Cluster^{3D}$ pair is used to reconstruct the merged $\pi^{0}$. A workflow chart of the merged $\pi^{0}$ reconstruction is presented in Fig.~\ref{fig:MergedPi0Rec}b. The following subsection focuses on the algorithms associated with the first and last steps in the workflow chart, which do not appear in single-photon reconstruction.

\subsubsection{Searching the second $Seed^{2D}$ in $Cluster^{2D}$s}
During the reconstruction process, a merged $\pi^{0}$ indicates that the generated photon pair was reconstructed as one $Cluster^{3D}$ in ECAL. This indicates that energy deposition from $\pi^{0}$ produces only one local maximum energy $Cell^{3D}$. For the unlayered-reconstruction framework in some experiments \cite{Machefert:691636, Terrier:691743} where layered readout information is lacking, it is common practice to select one of the non-seeded cells in the cluster with the highest energy cell as the second seed. A new cluster is then constructed around the second seed as the center. After splitting the energy of the shared cells between the new and original clusters, the overlapping clusters are referred to as subcluster pairs of the original cluster.

As shown in Fig.~\ref{fig:MergedPi0Rec}a, a merged $\pi^0$ does not necessarily result in "merged" $Cluster^{2D}$ in each layer. According to the cell-energy-splitting algorithm described above, the success rate of splitting increased as the number of cells shared by the two clusters decreased.

Furthermore, it is necessary to base the selection of the second $Seed^{2D}$ on the energy of $Cell^{2D}$s. This is because the energy of $Cell^{2D}$s are also related to the distance from the hit point of the photon. When one of the photons resulting from the decay of a $\pi^{0}$ has a much higher energy than the other photon in non-$Seed^{2D}$ $Cell^{2D}$s, the $Cell^{2D}$ closer to the higher-energy photon may have a higher energy than the $Cell^{2D}$ closest to the lower-energy photon. Solely focusing on the energy of $Cell^{2D}$ may result in misidentifying the second $Seed^{2D}$.

Hence, motivated by the aforementioned reasons, the layered reconstruction framework described in this study involves a two-step process for identifying the second $Seed^{2D}$ to enhance the accuracy of splitting $Cluster^{2D}$ and ultimately improve the efficiency of $\pi^{0}$ reconstruction. 

The first step is to search for $Cell^{2D}$, except for $Seed^{2D}$, which has the highest energy $E'$ in $Cluster^{2D}$ as $Cell^{2D'}$. The definition of $E'$ is given by the following equation:
\begin{equation}\label{eq:ETimesRatio}
E'= \frac{E(Cell^{2D})}{Frc},
\end{equation}
where $Frc$ is the estimated ratio of $Cell^{2D}$ obtained from the relationship shown in Fig.~\ref{fig:ESplitting}b. The second step is to determine whether there is any local energy maximum $Cell^{2D}$ other than $Seed^{2D}$ in the neighboring $Cell^{2D}$s of $Cell^{2D'}$ determined in the first step. If, then $Cell^{2D}$ with this local energy maximum is taken as the second $Seed^{2D}$; otherwise, $Cell^{2D'}$ is taken as the second $Seed^{2D}$. Subsequently, the second $Seed^{2D}$ is used to construct a new $Cluster^{2D}$ and energy splitting for the shared $Cell^{2D}$ and correction for $Cluster^{2D}$ are performed as detailed in the previous sections.

\subsubsection{$Cluster^{2D}$ matching}
After completing the preliminary algorithm, a sub-$Cluster^{2D}$ pair was obtained for each layer. The algorithm described in this section aims to match the sub-$Cluster^{2D}$s in different layers and finally obtain two sub-$Cluster^{3D}$s. The sub$Cluster^{3D}$ pair is regarded as a photon generated by the merged $\pi^0$. The matching of $Cluster^{2D}$ on different layers directly affects the accuracy of the reconstruction of the final sub-$Cluster^{3D}$. The utilization of multidimensional information from different readout layers facilitates a more accurate $Cluster^{2D}$ matching. 

First, the energy of the sub-$Cluster^{2D}$s in each layer was used for prematching. Because the ratio of the deposition energy in each layer is related to the energy of the incident particles\cite{livan2019}, unreasonable matches can be filtered based on the energy ratios between $Cluster^{2D}$s and the energy of $Cluster^{3D}$ as shown in Fig.~\ref{fig:CluFlow2}b. For example, in the case of a dual-layer ECAL, the energy of the $Cluster^{2D}$s in the front layer is used to calculate the energy of the $Cluster^{2D}$s in the other layers, and if the energy of the $Cluster^{2D}$s in the other layers exceeds the calculated value by $3\sigma$, the pre-matching is considered a failure, and the matching result is filtered.

If there is more than one $Cluster^{2D}$ in a certain layer that is prematched with the front layer, the final match is made based on the positional relationship. First, $Cluster^{2D}$ in the front layer is connected to the initial vertex (typically, the zero point). The connection line is extended and projected onto these layers, and the closest pre-matched $Cluster^{2D}$s to the projection point are matched with the front layer $Cluster^{2D}$ to complete the matching and obtain sub-$Cluster^{3D}$.

Finally, to avoid the resolution of $\pi^0$ being reconstructed in a merged model, $Seed^{2D}$s in all layers of sub-$Cluster^{3D}$ must be checked to determine whether they are included in the direct reconstruction $Cluster^{3D}$. If so, the current process of the merged $\pi^0$ reconstruction is terminated, and a new process starts by skipping the next directly reconstructed $Cluster^{3D}$.

\section{Performance}\label{sec.IV}

\begin{figure*}[!bht]
\includegraphics
  [width=0.98\hsize]
  {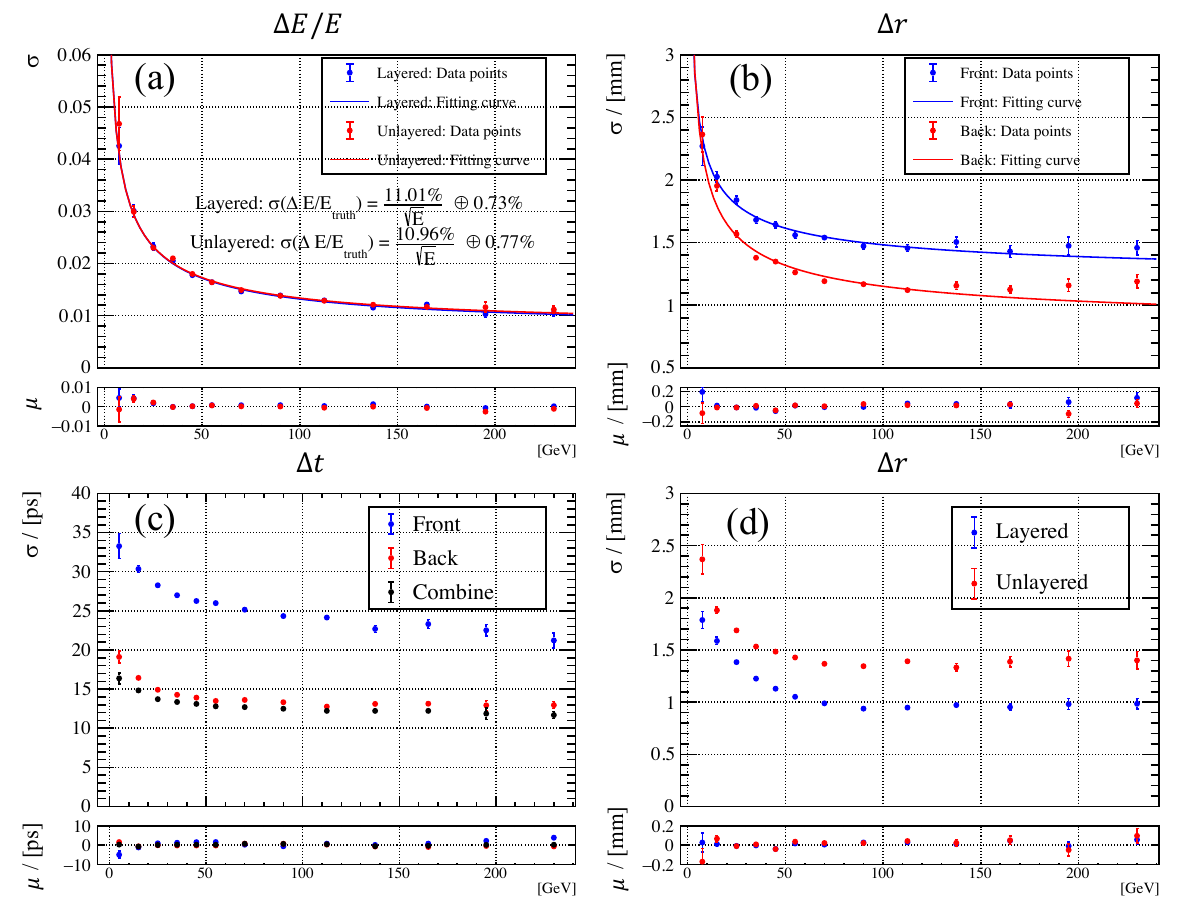}
\caption{The resolution and bias relative to the true value in regions 4-7. $\sigma$ and $\mu$ are derived from the Gaussian fitting, where $\sigma$ represents the standard deviation (used to denote the resolution) and $\mu$ is the mean value (used to denote the bias). (a): The energy resolution and bias, where $\Delta E$ is the difference between the true energy and the reconstructed energy, and $E$ is the true energy. (b): The position resolution and bias of front and back layers, where $\Delta r$ is the difference between the true position and the reconstructed position. (c): The time resolution and bias, where $\Delta t$ is the difference between the true time and the reconstructed time. (d): The position resolution and bias, where $\Delta r$ is the difference between the true position and the reconstructed position.}
\label{fig:Resolution}
\end{figure*}

\begin{figure}[!bht]
\includegraphics
  [width=0.98\hsize]
  {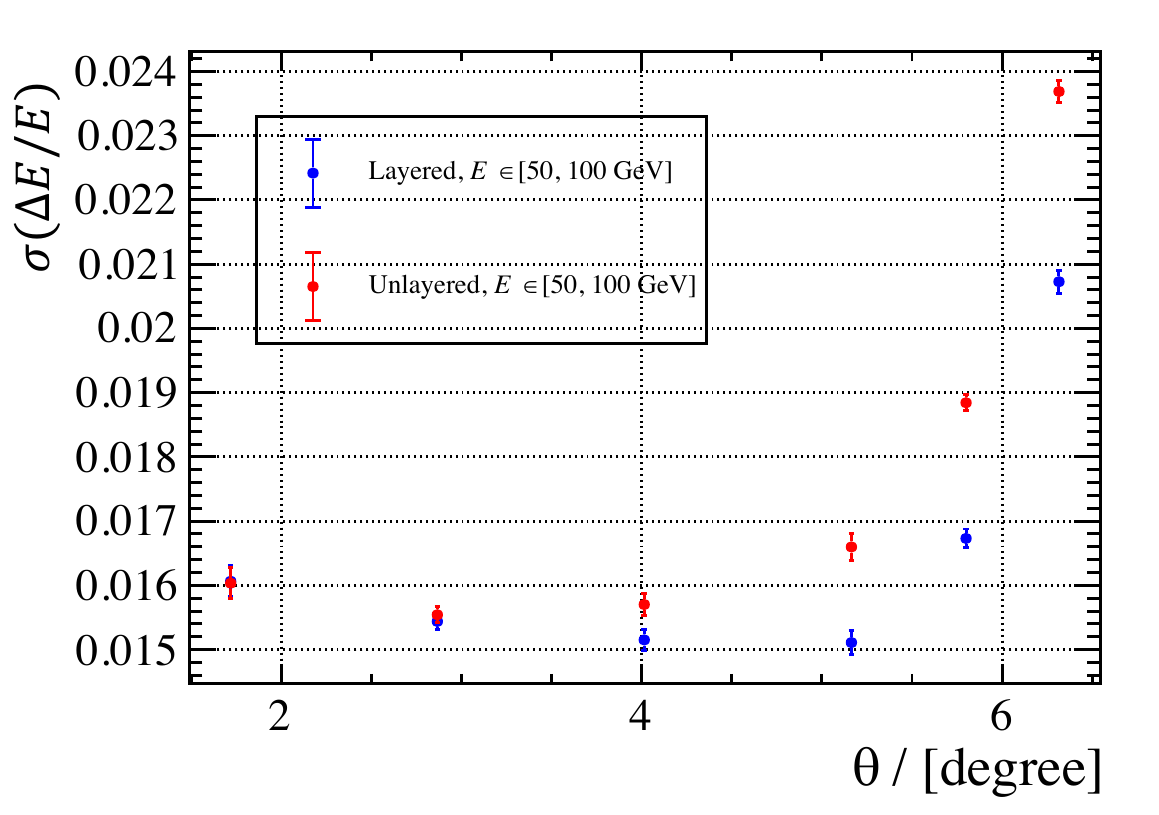}
\caption{The energy resolution versus angle $\theta$ in regions 4-7. Here, $\theta$ is the angle between the particle motion direction and beam direction. $\Delta E$ represents the difference between the true and reconstructed energies, and $E$ represents the true energy. $\sigma$ is the standard deviation derived from the Gaussian fitting. }
\label{fig:ERes_VsAngle}
\end{figure}

\begin{figure*}[!bht]
\includegraphics
  [width=0.96\hsize]
  {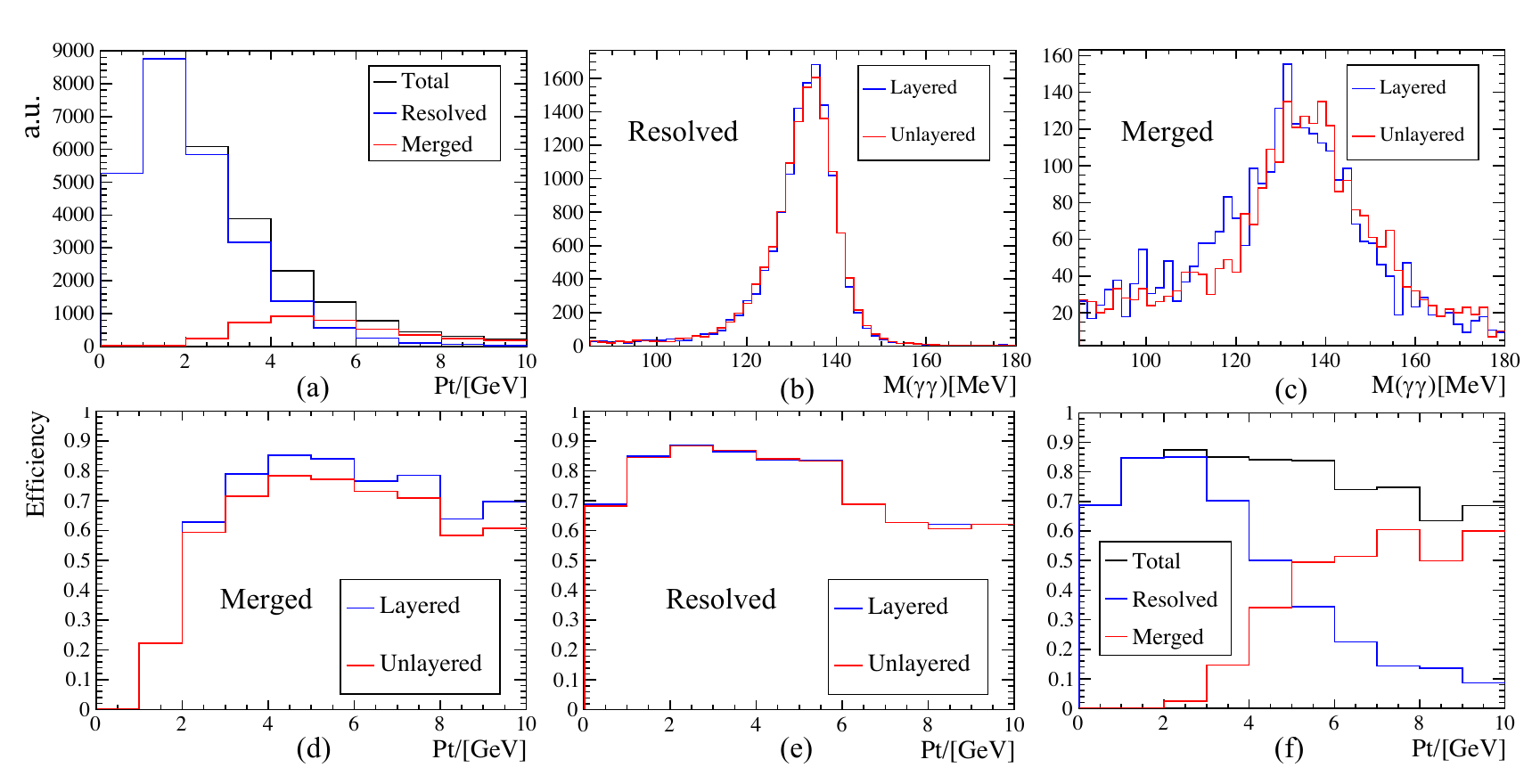}
\caption{(a): The transverse momentum distribution of $\pi^{0}$ from $B^0\rightarrow\pi^+\pi^-\pi^0$. (b): The distribution of $M(\gamma\gamma)$ of matched candidates in the resolved model in the SpaCal region. (c): The distribution of $M(\gamma\gamma)$ of matched candidates in the merged model in the SpaCal region. (d):  The reconstruction efficiency of merged $\pi^{0}$ from $B^0\rightarrow\pi^+\pi^-\pi^0$. (e): The reconstruction efficiency of resolved $\pi^{0}$ from $B^0\rightarrow\pi^+\pi^-\pi^0$.  (f): The total reconstruction efficiency of $\pi^{0}$ from $B^0\rightarrow\pi^+\pi^-\pi^0$ in layered reconstruction framework.}
\label{fig:Pi0_Result}
\end{figure*}

In the context of the LHCb experiment, a Phase-\Rmnum{2} upgrade (LHCb Upgrade \Rmnum{2}) was proposed\cite{The_LHCb_Collaboration_2008}. Scheduled for installation at the beginning of LHC Run 5 around 2036, the LHCb Upgrade \Rmnum{2} aims to enhance the experimental capabilities for exploring the frontiers of particle physics. PicoCal was designed with a longitudinally layered ECAL structure. The Shashlik calorimeter structure \cite{Amato:494264,Barsuk:691508} was retained in the outer region of PicoCal, whereas the Spaghetti calorimeter (SpaCal) \cite{Jenni:181281,AN2023167629} was used in the central region. The GAGG crystal, which is known for its high radiation resistance, high light yield, and excellent time-response performance\cite{ALENKOV2019226,doi:10.1021/cg200694a,doi:10.1021/cg501005s,LUCCHINI2016176,MARTINAZZOLI2021165231,7410106}, was introduced as a sensitive material in the most central region with the highest radiation dose. Reduced detector occupancy can be achieved by designing and using modules with smaller Molière radii to achieve smaller readout cell sizes in the internal regions with the highest detector occupancies. The detailed layout is presented in Table~\ref{tab:Modules}. 

In this section, based on the above layout, a series of single-photon and $\pi^0$ samples are used to demonstrate the performance of the layer-reconstruction algorithm in this framework. In addition, the unlayered-reconstruction algorithm\cite{LHCb-2003-091} employed in Run 1/2 of LHCb was introduced for comparison. To ensure fairness, the unlayered-reconstruction algorithm remained consistent with the layered reconstruction algorithm in terms of cluster correction, overlapping cluster splitting, and other steps, except for the utilization of layered information.

\subsection{Single-photon performance}
The single-photon samples utilized in this section were generated and simulated using the "Hybrid MC" simulation framework \cite{SimCode}, which is built upon the GEANT4 Monte Carlo package \cite{AGOSTINELLI2003250}. The performance of the layered reconstruction algorithm in this framework was demonstrated in terms of the energy, position, and time resolution.

\begin{table}[!htb]
\caption{Modules in 2023 baseline setup of PicoCal.}
\label{tab:Modules}
\begin{tabular*}{8.6cm} {@{\extracolsep{\fill} } cccccc}
\toprule
\thead{Region} & \thead{Type} & \thead{Absorber/Crystal} &\thead{Cell size \\ $\left[\mathrm{cm}^2\right]$} & \thead{$R_M$ \\$\left[\mathrm{mm}\right]$} & \thead{Layers}\\
\midrule
1  & Shashlik & Lead/Polystyrene & 12$\times$12 &35.0 & 2\\
2   & Shashlik & Lead/Polystyrene & 6$\times$6 &35.0 & 2\\
3   & Shashlik & Lead/Polystyrene & 4$\times$4 &35.0 & 2\\
4-7   & SpaCal & Lead/Polystyrene &3$\times$3 &29.5 & 2\\
8-11 & SpaCal & Tungsten/GAGG &1.5$\times$1.5 &14.5 & 2\\
\bottomrule
\end{tabular*}
\end{table}

\subsubsection{Energy resolution}
The Fig.~\ref{fig:Resolution} (a) illustrates the energy resolution versus the incident energy. This relationship can be described by the following equation:
\begin{equation}\label{eq:Resolution}
\sigma(E)/E = \frac{a}{\sqrt{E}} \oplus b,
\end{equation}
where $a$ represents the statistical fluctuations of the readout signal (e.g., photoelectrons), $b$ represents the constant term errors owing to uncertainties in the calibration and correction, as well as inhomogeneities in the active material, and $\oplus$ represents the sum of squares. The energy resolution and bias as functions of energy illustrate the stability of the layered reconstruction framework. In addition, as shown in Figs.~\ref{fig:ERes_VsAngle}, the layered reconstruction algorithm demonstrates a better energy resolution for particles with large incident angles. However, because of the forward detector design and layout of PicoCal\cite{Scoping_Document}, there are fewer single-photon events with large incident angles. Therefore, this improvement was not significantly reflected in the energy resolution as a function of the incident particle energy.

\subsubsection{Position resolution}
The position resolution is a critical parameter of the ECAL and is, to some extent, equivalent to the angular resolution. The positional resolutions are shown in Fig.~\ref{fig:Resolution}b and d, respectively, where $Front$ and $Back$ represent the front and back layers, respectively, $r$ represents the position. The development stages of the shower and readout layers change as the particles undergo showering and deposit energy along their momentum direction. However, the tendency of the reconstructed $Cluster^{2D}$'s raw position, relative to the true position, varied between the different layers. Essentially, if we only have the energy-weighted position information of $Cell^{2D}$s located in different layers, the raw position will spread out in the transverse plane owing to different tendencies. Consequently, utilizing an unlayered-reconstruction algorithm and an overall correction parameter for reconstructing the transverse position will lead to degradation of the position resolution owing to this spreading effect. In contrast, the reconstruction in this study effectively resolved the previously mentioned issue and improved the positional resolution, as illustrated in Fig.~\ref{fig:Resolution}d. This enhancement results in an improvement of approximately 0.5 mm in the high-energy region, representing a 33\% increase compared with the unlayered-reconstruction algorithm.

\subsubsection{Time resolution}
Time is essential for event reconstruction and data analysis in high-luminosity environments. Within the layered reconstruction framework, time information can be provided in the form of $Cluster^{3D}$ or along the longitudinal direction using $Cluster^{2D}$, as shown in Fig.~\ref{fig:Resolution}c. The integration of layered time information provides new analytical perspectives for future physical investigations. Ongoing research also focuses on exploring the application of time information for reconstruction, correction, and analysis.

\subsection{$\pi^{0}$ reconstruction performance}
In this section, approximately 30,000 signal $\pi^0$ events from $B^0\rightarrow\pi^+\pi^-\pi^0$ are generated by the generation part of Gauss\cite{Clemencic_2011}, with the requirement that all final-state photons from $\pi^0$ should be contained within the acceptance region of the ECAL. The transverse momentum distribution of $\pi^0$ is shown in Fig.~\ref{fig:Pi0_Result}a. These samples were used as benchmarks to test the contribution of the layered reconstruction algorithm in this framework to the reconstruction performance of $\pi^0$ particles. The matched $M(\gamma\gamma)$ distribution, which was compared with the unlayered-reconstruction algorithm, is shown in Fig.~\ref{fig:Pi0_Result}b and Fig.~\ref{fig:Pi0_Result}c. The reconstruction efficiency of $\pi^0$ based on the layered reconstruction framework is shown in Fig.~\ref{fig:Pi0_Result}f. An efficiency comparison of the layered and unlayered-reconstruction algorithms is presented in Fig.~\ref{fig:Pi0_Result}d and Fig.~\ref{fig:Pi0_Result}e, where the resolved and merged modes are presented separately.

As shown in Fig.~\ref{fig:Pi0_Result}d, as expected, utilizing the layered reconstruction algorithm in this framework improves the ability to split overlapping $Cluster^{3D}$ s, leading to a 10$\%$ increase in the efficiency of reconstructing the merged $\pi^0$. The improvement in the position resolution of a single-photon contributes to the improvement in the resolution of $\pi^0$ mass distribution, as shown in Fig.~\ref{fig:Pi0_Result}b and Fig.~\ref{fig:Pi0_Result}c.

\begin{figure}[!bht]
\includegraphics
  [width=0.98\hsize]
  {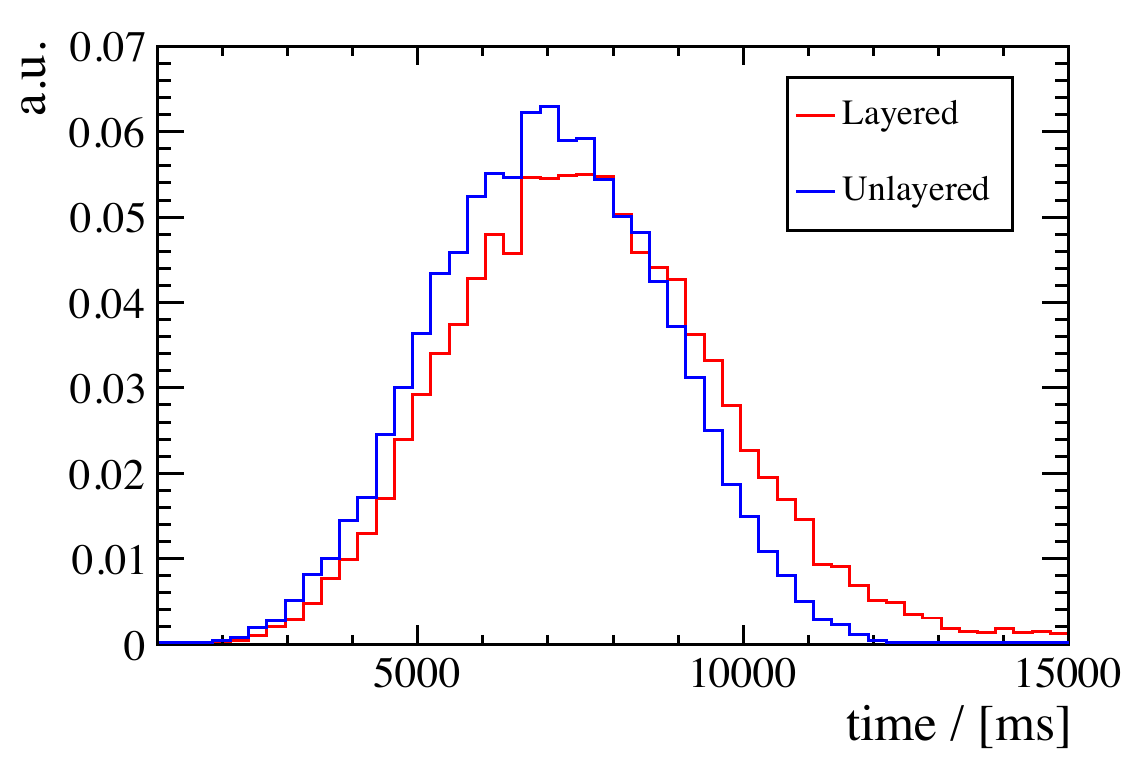}
\caption{The time consumption distribution for each event.}
\label{fig:ComputeTime}
\end{figure}

\begin{table*}[!htb]
\caption{The time complexity.}
\label{tab:Time_complexity}
\begin{tabular*}{10cm} {@{\extracolsep{\fill} } cc}
\toprule
Algorithm & Time complexity \\
\midrule
Construct $Cell^{3D}$s & $O(n_1)$\\
Search $Seed^{3D}$s&$O(n_2)$ \\
Construct $Cluster^{3D}$s&$O(n_3)$ \\
Split overlapping $Cluster^{3D}$s&$O(n_{4}^2)$ \\
Search $Seed^{2D}$s/Construct $Cluster^{2D}$s& $O(n_4)$   \\
Correct $Cluster^{3D}$s & $O(n_4)$ \\
\bottomrule
\end{tabular*}
\end{table*}

\section{Computation}\label{sec.V}
This framework allows task splitting according to events, thereby enabling the deployment of different events as separate tasks. Based on this framework, we evaluated the runtime of $Cluster^{3D}$ reconstruction under LHCb PicoCal at a center-of-mass energy of 14 TeV and an instantaneous luminosity of $1.5\times10^{34}\  \text{cm}^{-2}\text{s}^{-1}$. The framework was deployed on a cluster CPU with an $Intel(R)\ Xeon(R)\ Platinum\ 9242\ CPU\ @\ 2.30\ GHz$ for each task. Because events are independent of each other in this framework, we considered only the time required for the reconstruction of each individual event. For comparison, we also introduced the unlayered-reconstruction algorithm used in LHCb, and the results are shown in Fig.~\ref{fig:ComputeTime}. Compared with the unlayered-reconstruction algorithm, the layered reconstruction algorithm did not significantly increase the computation time.

Scalability is a critical consideration for CPU cluster design. In this framework, an event is defined as the smallest unit of a cluster task. This approach facilitated efficient task management and resource allocation. By increasing the number of computing nodes, the number of events processed in parallel can be increased, thereby reducing the overall task runtime. The computational speed of event reconstruction can be enhanced by using more powerful CPUs in each computing node. 

The time complexity of each step in the reconstruction process for each event is shown in the flowchart in Fig.~\ref{fig:CluFlow2}, and are listed in Table~\ref{tab:Time_complexity}. Here, $n_1$ represents the number of $Cell^{2D}$s, $n_2$ represents the number of $Cell^{3D}$s, $n_3$ represents the number of $Seed^{3D}$s, and $n_4$ represents the number of $Cluster^{3D}$s. Considering the parallelism within an event, the structure of these processes in this framework was designed for future deployment on nodes with parallel computing capabilities, such as GPUs or FPGAs. This establishes a solid foundation for the future deployment and acceleration of these algorithms in GPU clusters and FPGA platforms.

\section{Conclusion}\label{sec.VI}
As depicted in Fig.~\ref{fig:Reconstruction_Framework}, this work has accomplished we developed a software framework for the longitudinally segmented ECAL event reconstruction. Moreover, a layered reconstruction algorithm was devised within this framework for $Cluster^{3D}$s and the merged $\pi^0$. This framework not only furnishes the general direction, arrival time, and energy of the particle candidates in the $Cluster^{3D}$ format, but also provides the position, timestamp, and energy deposition in each layer in the $Cluster^{2D}$ format. The information from the $Cluster^{2D}$s can not only be used to filter out unreasonable $Cluster^{3D}$s during reconstruction, but also to provide new perspectives in physics analysis.

To achieve a more refined correction of the $Cluster^{3D}$ information, this study leveraged the advantages of a layered reconstruction framework to provide layered information and designed a layered correction method and process. In terms of energy correction, compared with solely using the $Cluster^{3D}$ energy for correction, this study further utilized the energy ratios of $Cluster^{2D}$ in each layer for correction, aiming to better compensate for the longitudinal variation in the sampling fraction of the ECAL. To correct the time and position information, we first corrected the corresponding information of the $Cluster^{2D}$s and then weighted the corrected $Cluster^{2D}$ information based on the resolution of the corresponding information in each layer. The weighted result was used as the corrected information for $Cluster^{3D}$. Moreover, this study delves into the transverse shower profile and systematically elucidates the relationship between the distance and energy ratio between $Cell^{2D}$ s and $Cluster^{2D}$. This information provides more precise prior knowledge for splitting overlapping $Cluster^{3D}$s.

Finally, the performance of the framework was validated using the PicoCal in LHCb Upgrade \Rmnum{2}. The results show that the layered reconstruction algorithm in this framework significantly improves the position resolution of a single photon, and the energy resolution of the particles at large incident angles compared with the unlayered-reconstruction algorithm. For example, in regions 4-7 of the specified setup, the position resolution was enhanced from approximately 1.4 mm to 0.9 mm in the high-energy region. In addition, the energy resolution was improved by approximately 10\% at large incident angles. Furthermore, the layered reconstruction algorithm enhances the splitting capability of overlapping clusters, leading to further improvements in the efficiency of merged $\pi^0$ reconstruction. The current version of the algorithm can increase the reconstruction efficiency of the merged $\pi^0$ by approximately 10\% for SpaCal in the aforementioned setup.

Furthermore, this study provided a suitable software platform for future studies on layered ECAL. It incorporates comprehensive data structures and application programming interfaces (APIs), along with straightforward configuration and execution. This feature allows convenient secondary development by leveraging the framework to substitute and validate new algorithms. This also facilitates the investigation of ECAL-related physics. In future work, we will continue to explore various facets by utilizing the multidimensional information and scalability provided by the proposed software framework. This includes delving into the application of deep learning in cluster splitting and information correction, evaluating the performance of different cluster shapes, and scrutinizing the application of time information in cluster reconstruction.

\section{Acknowledgments}\label{sec.VII}
We express our gratitude to Marco Pizzichemi, Loris Martinazzoli, and Philipp Gerhard Roloff of LHCb ECAL Upgrade \Rmnum{2} R\&D group for developing the simulation software and for their helpful suggestions. Additionally, we extend our thanks to the developers and maintainers of the Gauss software in LHCb, which we used to generate events. The numerical calculations in this study were performed using a supercomputing system at the Supercomputing Center of Wuhan University.

\end{document}